\documentclass[twoside,fleqn]{article}
\usepackage{epsf}
\usepackage{espcrc2}

\newcommand{\pspicture}[1]{
\setlength\epsfysize{180pt}
\setlength\epsfxsize{180pt}\epsfbox{#1}}

\title{
Current Renormalisation in NRQCD
for Semi-leptonic $B\rightarrow D$ Decays}
\author{Peter Boyle, Christine Davies
\address{Dept. of Physics and Astronomy, 
University of Glasgow, Glasgow, UK ; UKQCD Collaboration
}
}

\begin{document}

\begin{abstract}
We present a calculation of the renormalisation constants for 
the temporal vector current, $Z_{V_0}$
and spatial axial current, $Z_{A_k}$, to $O(\frac{\alpha_s}{M})$
for $B\rightarrow D$ transitions using the $O(\frac{1}{M})$ NRQCD action
for both $b$ and $c$ quarks evaluated for a large range of mass parameters.
Considerations for the renormalisation of the spatial vector current
and the temporal axial current are discussed and initial results
for a mixed lattice current are presented for the spatial vector current.
\end{abstract}

\maketitle

\section{Introduction}
\vspace{-.05in}
The semi-leptonic $B\rightarrow D,D^*$ decay is phenomenologically interesting
since its Feynman amplitude involves the poorly know CKM matrix
element $V_{cb}$
\begin{equation}
M(B\rightarrow X l^- \bar{\nu}) = -\frac{i}{2} V_{cb} \bar{u}_l \gamma^\mu (1-\gamma_5) v_\nu H_\mu
\end{equation}
\begin{equation}
H_\mu = \langle X(p^\prime) | \bar{c}\gamma_\mu (1-\gamma_5) b | B(p)\rangle~~~,~~~X=D,D^*
\end{equation}
where the hadronic tensor $H_\mu$, and its Lorentz decompostion into
form factors is calculable using lattice QCD.

We perform the matching of the Lattice to continuum $\overline{MS}$  currents 
to 1 loop using the on-shell scheme in Feynman gauge with
vanishing external spatial momenta. In the on-shell
scheme the wavefunction renormalisation and vertex corrections
individually contain (cancelling) infra-red divergences 
which we control using a fictitious gluon mass, $\lambda$.
We use dimensional regularisation to control the ultra-violet 
behaviour in the continuum calculation.

\vspace{-.05in}
\section{Continuum Calculation}
\vspace{-.05in}
Expanding the 1-loop correction to the currents to $O(\frac{\vec{p}}{M})$,
and writing $\epsilon = \frac{M_c}{M_b}$, we obtain
\begin{equation}
\delta V_0 / \frac{\alpha_s}{3\pi} =  \gamma_0 \left[ 3\frac{\epsilon+1}{\epsilon-1} \log{ \epsilon} - 4 \right]
          - 2  + O(\frac{1}{M^2})
\end{equation}

\begin{equation}
\label{ContinuumVk}
\begin{array}{lll}
\delta V_k /\frac{\alpha_s}{3\pi} &=& \gamma_k \left[ 3\frac{\epsilon+1}{\epsilon-1} \log{ \epsilon} - 4 \right]\\
&          +& \frac{p_k}{M_b} \left[\frac{2}{\epsilon-1} - \frac{2 \epsilon \log\epsilon}{(\epsilon-1)^2}\right] \\
&          +& \frac{p^\prime_k}{M_c} \left[\frac{2 \epsilon}{\epsilon-1} +\frac{2 \epsilon \log\epsilon}{(\epsilon-1)^2}\right]+ O(\frac{1}{M^2})
\end{array}
\end{equation}

\begin{equation}
\delta A_k /\frac{\alpha_s}{3\pi} = \gamma_k\gamma_5 \left[ 3\frac{\epsilon+1}{\epsilon-1}\log\epsilon - 8 \right] + O(\frac{1}{M^2})
\end{equation}

\begin{equation}
\begin{array}{lll}
\delta A_0 /\frac{\alpha_s}{3\pi} &=& \gamma_0\gamma_5 \left[ 3\frac{\epsilon+1}{\epsilon-1}\log\epsilon - 8 \right]\\
& + &\gamma_5 \left[ 6\frac{\epsilon+1}{\epsilon-1}\right]
 + O(\frac{1}{M^2})
\end{array}
\end{equation}

Here the currents $V_0$ and $A_k$ can be renormalised to $O(\frac{\vec{p}}{m})$
in the usual manner, however the currents $V_k$ and $A_0$ will involve
operator mixing at this order.

\vspace{-.05in}
\section{1-loop Lattice Correction}
\vspace{-.05in}
We use the $O(\frac{1}{M})$ lattice NRQCD action \cite{LepageNRQCD} 
for both $b$ and $c$ quarks,
\begin{equation}
\begin{array}{l}
a{\cal L}_{\rm NRQCD} =  \psi^\dagger(x) \psi(x) \\
-	
\psi^\dagger({\small x+\hat{t}})
\left( 1-\frac{a\delta H}{2} \right)
\left( 1 - \frac{a H_0}{2 n}\right)^n
\frac{U_4^\dagger(x)}{u_0} \\
\times
\left( 1 - \frac{a H_0}{2 n}\right)^n
\left( 1-\frac{a\delta H}{2} \right)
\psi({\small x})
\end{array},
\end{equation}
\begin{equation}
H_0 = -\frac{\Delta}{2M_0} ~~~~,~~~~\delta H = - C_B \frac{g {\bf \sigma}\cdot {\bf B}}{2M_0}
\end{equation}

for which the Feynman rules may be found in reference 
\cite{JunkColinJournalPaper}. 

The Foldy-Wouthysen transformation of the continuum current 
corrections and dimensional arguments for which currents can contribute
at $O(\frac{1}{M})$ suggest we take the bases of operators in Table~\ref{TabCurrents}
for the lattice currents.
\begin{table}[bt]
\caption{Lattice Current Bases}
\label{TabCurrents}
\begin{tabular}{cccc}
\hline
$V_0$ & $V_k$ & $A_0$ &$A_k$ \\
\hline
$ 1 $
& 
$\begin{array}{l}
J^{V_k}_1 = - \sigma_k \frac{\sigma \cdot \stackrel{\rightarrow}{\nabla} }{2M_b} \\
J^{V_k}_2 = \frac{\sigma \cdot \stackrel{\leftarrow}{\nabla} }{2M_c} \sigma_k \\
J^{V_k}_3 = - \frac{ \stackrel{\rightarrow}{\nabla}_k }{M_b} \\
J^{V_k}_4 = \frac{ \stackrel{\leftarrow}{\nabla}_k }{M_c} 
\end{array} $ 
&
$
\begin{array}{c}
J^{A_0}_1 = - \frac{\sigma \cdot \stackrel{\rightarrow}{\nabla} }{2M_b} \\
J^{A_0}_2 =  \frac{\sigma \cdot \stackrel{\leftarrow}{\nabla} }{2M_c} 
\end{array} 
$
&
$ \sigma_k$\\
\hline
\end{tabular}
\vspace{-.32in}
\end{table}

For those currents that are unmixed, the calculation is similar to \cite{CTHD_BETH}, and we
write the renormalisation constant $Z_{\Gamma}  =  1 + \alpha_s Z_{\Gamma}^{[1]}
= 1 + \delta Z_\Gamma^{\rm \overline{MS}} - \delta Z_\Gamma^{\rm lat}$, where the 
lattice integrals for 
$\delta Z_\Gamma^{\rm lat}$ were
computed numerically using VEGAS \cite{LepageVEGAS},
performing the temporal loop momentum integration analytically for those contributions
that were infra-red divergent.

In the case of the spatial vector and temporal axial currents
it is necessary to evaluate the derivatives of the diagrams 
with respect to the external momentum to match with the
continuum. This is done by numerically evaluating the integral of the analytically taken
derivative, resolving the pieces of the Pauli structure
by taking different derivative directions.

In the continuum, both the vertex correction to the 
current $\bar{c}\Gamma b$, and the wavefunction
renormalisation contained a logarithmic divergence proportional to 
$\frac{4}{3\pi} \log \frac{\lambda}{M} \Gamma $ where the sign is positive
in the vertex correction. The same infra-red divergences were
found in the lattice vertex correction and wavefunction renormalisation
so that as one would expect the infra-red behaviour of the theory is identical
to that in the continuum.

\vspace{-.07in}
\subsection {$V_0$ and $A_k$ Currents}
\begin{figure}[hbt]
\pspicture{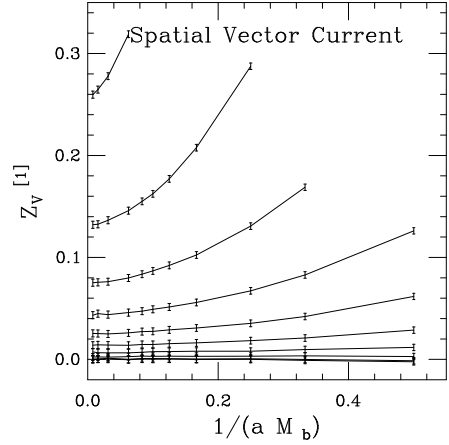}
\vspace{-.3in}
\caption{$Z_{V_0}^{[1]}$ as a function of $\frac{1}{M_b}$ and $\frac{M_c}{M_b}$.
The lines correspond to $\frac{M_c}{M_b} = 0.1, \ldots, 1.0$ in increments of
$0.1$, with the $\frac{M_c}{M_b} = 0.1$ curve topmost.}
\label{FigZV0}
\vspace{-.3in}
\end{figure}

The 1-loop contribution to the renormalisation constant $Z_{V_0}$
is plotted in Figure \ref{FigZV0} for various values of $\epsilon = \frac{M_c}{M_b}$ and $M_b$.
Here the stabilisation parameter $n$ has been chosen in a mass dependent manner
such that $M_0 \times n \ge 3$.
The lattice current is the conserved current of the lattice action,
and the continuum current is conserved, so that there is no correction on the
line $\epsilon = 1$.

\begin{figure}[hbt]
\pspicture{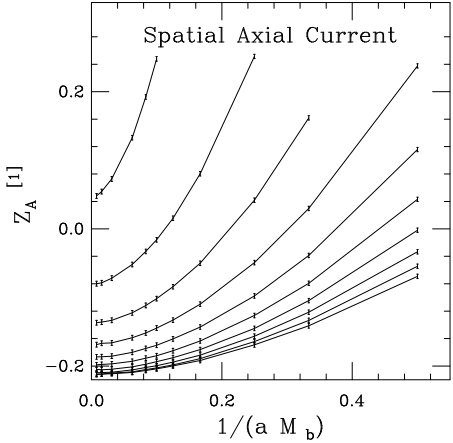}
\vspace{-.3in}
\caption{$Z_{A_k}^{[1]}$ as a function of $\frac{1}{M_b}$ and $\frac{M_c}{M_b}$.
The lines correspond to $\frac{M_c}{M_b} = 0.1, \ldots, 1.0$ in increments of
$0.1$, with the $\frac{M_c}{M_b} = 0.1$ curve topmost.}
\label{FigZAk}
\vspace{-.3in}
\end{figure}

The 1-loop correction to the spatial axial current is in 
Figure \ref{FigZAk}. The mass dependence is stronger
since the line for $\epsilon=1$ is not protected.

\vspace{-.07in}
\subsection {$V_k$ Current}
We define the 1-loop lattice mixing matrix $Z_{ij}$,
\begin{equation}
\langle c(p^\prime) | J_{i} | b(p) \rangle = \sum\limits_j Z_{ij} \Omega^E_j
\end{equation}
where the $\Omega^E_j$ are the continuum analogs of the lattice operators
given in Table \ref{TabCurrents}.
We define the contribution to the mixing matrix arising from the lattice 
vertex correction, $\tilde{\xi}_{ij}$, via
\begin{equation}
Z_{ij} = \delta_{ij} + \alpha_s \left[
\begin{array}{c} 
\frac{1}{2} \{ \tilde{Z}_{\psi_c}^{\rm lat} 
+ \tilde{Z}_{\psi_b}^{\rm lat} \} \delta_{ij} 
+ \tilde{\xi}_{ij} \\
+ \tilde{Z}_{m_b}^{\rm lat} \delta_{i1}\delta_{j1}
+ \tilde{Z}_{m_c}^{\rm lat} \delta_{i2}\delta_{j2}
\end{array}
\right]           
\end{equation}
We then invert
the mixing matrix and match to the continuum.
Here the coefficients $B_i$ may be inferred from equation \ref{ContinuumVk}, and
the $Z_m$ factors arise from matching the bare mass in lattice currents 
to the pole mass in the tree level continuum current.

\begin{eqnarray}
\label{EqnMixingBare}
\nonumber
V_k^{\rm \overline{MS}} &=&\left[ 
1 + \alpha_s \left(
\begin{array}{c}
B_1 -\frac{1}{2} \{ \tilde{Z}_{\psi_c}^{\rm lat}+\tilde{Z}_{\psi_b}^{\rm lat} \} \\
 - \tilde{\xi}_{11} - \tilde{\xi}_{21} - \tilde{Z}_{m_b}
\end{array}
 \right) \right] J_1 \\
\nonumber
&+& \left[ 1 + \alpha_s \left( 
\begin{array}{c}
B_2 -\frac{1}{2} \{ \tilde{Z}_{\psi_c}^{\rm lat}+\tilde{Z}_{\psi_b}^{\rm lat} \} \\
 - \tilde{\xi}_{12} -\tilde{\xi}_{22} -\tilde{Z}_{m_c}
\end{array}
\right) \right] J_2 \\
\nonumber
&+& ~~~~~~~ \alpha_s \left( B_3 - \tilde{\xi}_{13} - \tilde{\xi}_{23} \right) J_3 \\
&+& ~~~~~~~ \alpha_s \left( B_4 - \tilde{\xi}_{14} - \tilde{\xi}_{24} \right) J_4 
\end{eqnarray}

The diagonal elements $\tilde{\xi}_{jj}$ contain infra-red divergences which cancel
with those in the lattice wavefunction renormalisations.
Those currents appearing at tree level also carry tadpole
improvement counter terms.
We therefore rewrite Equation~\ref{EqnMixingBare} in terms of 
IR divergence and tadpole counter term free quantities,
as follows: 
\begin{eqnarray}
\nonumber
V_k^{\rm \overline{MS}}
& = & 
\left[ 1 + \alpha_s \left( 
\begin{array}{c}
B_1 -\frac{1}{2} \{ Z_{\psi_c}^{\rm lat}+Z_{\psi_b}^{\rm lat} \}\\
-Z_{m_b} - \xi_{11} - \xi_{21} \\ -\xi_{11}^{\rm TI} -  Z_{m_b}^{\rm TI}
\end{array}
\right) \right] J_1\\
\nonumber
    &+& \left[ 1 + \alpha_s \left(
\begin{array}{c}
B_2 -\frac{1}{2} \{ Z_{\psi_c}^{\rm lat}+Z_{\psi_b}^{\rm lat} \}\\
-Z_{m_c} - \xi_{12} - \xi_{22} \\
-\xi_{22}^{\rm TI} - Z_{m_c}^{\rm TI}
\end{array}
\right) \right] J_2 \\
\nonumber
&+& \alpha_s \left( B_3 - \xi_{13} - \xi_{23} \right) J_3 \\
&+& \alpha_s \left( B_4 - \xi_{14} - \xi_{24} \right) J_4
\end{eqnarray}

Preliminary calculations of the mixed lattice have been performed on a few mass values
for degenerate quarks. 
The degenerate case may in fact be of use in certain lattice simulations to obtain
the Isgur-Wise function,
however the results should really be considered illustrative 
of the method, and the calculation over a similar parameter regime
to our previous results for the other currents will be performed.
In this case the coefficients of $J_1$ and
$J_2$, and of $J_3$ and $J_4$ are identical, and we only present the coefficients
for $J_1$ and $J_3$ in Table~\ref{TabOneLCoeff}. We evaluate the tadpole improvement
counter terms using $u_0^{[1]}({\rm plaq}) = \frac{\pi}{3}$, and
take values for $Z_m$ from \cite{JunkColinJournalPaper}. 
Also, for the degenerate case
$B_1 = B_2 = -\frac{2}{3\pi}$ and
$B_3 = B_4 = -\frac{1}{3\pi}$, so that when the mixed current is written as
\begin{eqnarray}
\nonumber
V_k^{\rm \overline{MS}} &=& \left( 1+\alpha_s C_1 \right) \left( J_1 + J_2 \right)\\
                   &+& \alpha_s C_2 \left( J_3 + J_4 \right) ,
\end{eqnarray}
we obtain the coefficients in Table~\ref{TabMixedCurrent}.

\begin{table}[bt]
\caption{One Loop Correction Coefficients}
\label{TabOneLCoeff}
\begin{tabular}{cccccc}
\hline
$M_Q$ & $\xi_{11}+\xi_{21}$ & $\xi_{13} + \xi_{23}$ & $Z_\psi$ \\
\hline
2 & -0.6581(5) &  -0.2516(1) & -0.4156(6) \\
4 & -0.9219(5) &  -0.2041(1) & -0.0030(6) \\
10& -1.2662(5) &  -0.0963(1) & -0.2711(8) \\
\hline
\end{tabular}
\vspace{-0.3in}
\end{table}

\begin{table}[bt]
\caption{One Loop Correction Coefficients}
\label{TabMixedCurrent}
\begin{tabular}{ccccccc}
\hline
$M_Q$,$n$ & $C_1$ & $C_2$ \\
\hline
2,2 & -0.776(4) & 0.145(1)\\
4,2 & -0.571(4) & 0.098(1)\\
10,1& -0.247(4) & -0.010(1)\\
\hline
\end{tabular}
\vspace{-0.2in}
\end{table}

\vspace{-.05in}
\section{Future Direction}
\vspace{-.05in}
We are nearing completion of the calculation of the mixed current for $V_k$,
while there is a smaller basis of operators for $A_0$, so that the corresponding
calculation should be somewhat easier. Thereafter we shall perform a similar calculation
for the $b\rightarrow c$ transition using the NRQCD action for the $b$ quark, and
the O(a) improved Wilson action for the $c$ quark.
The extension of the calculation to $O(\frac{\alpha}{M^2})$ would be of interest.

\vspace{-.05in}
\section{Acknowledgements}
\vspace{-.05in}
We would like to thank the Physics Department and the ITP, UCSB for their hospitality
while this work was carried out.
PB is funded by PPARC grant PP/CBA/62, CD thanks the Fulbright commission
and the Leverhulme trust.

\end{document}